\documentclass{aa}  

\usepackage{graphicx}
\usepackage{txfonts}

\usepackage{caption}
\usepackage{xcolor}
\usepackage[export]{adjustbox}
\usepackage{comment}

\begin{document}

   \title{Deep view of the intracluster light in the Coma cluster of galaxies}

   \titlerunning{ICL in the Coma cluster}
   \authorrunning{Jiménez-Teja et al. }

   \author{Yolanda Jim\'enez-Teja\inst{1,2}\thanks{   \email{yojite@iaa.es}}
   \and
   Javier Román \inst{3,4,5}
   \and
   Kim HyeongHan \inst{6}
   \and 
   Jose M. Vi\'lchez \inst{1}
   \and
   Renato A. Dupke \inst{2,7,8}
   \and 
   Paulo Afr\^anio Augusto Lopes \inst{9}
   \and
   Robert Michael Rich \inst{10}
   \and 
   Osmin Caceres \inst{10}\thanks{McNair Fellow}
   \and 
   Chester Li \inst{10,11}
   }

   \institute{Instituto de Astrof\'isica de Andaluc\'ia--CSIC, Glorieta de la Astronom\'ia s/n, E--18008 Granada, Spain
   \and Observat\'orio Nacional, Rua General Jos\'e Cristino, 77 - Bairro Imperial de S\~ao Crist\'ov\~ao, Rio de Janeiro, 20921-400, Brazil
   \and Departamento de Física de la Tierra y Astrofísica, Universidad Complutense de Madrid, E-28040 Madrid, Spain
   \and Instituto de Astrof\'{\i}sica de Canarias, c/ V\'{\i}a L\'actea s/n, E-38205, La Laguna, Tenerife, Spain
   \and Departamento de Astrof\'{\i}sica, Universidad de La Laguna, E-38206, La Laguna, Tenerife, Spain
   \and Department of Astronomy, Yonsei University, 50 Yonsei-ro, Seoul 03722, Korea.
   \and Department of Astronomy, University of Michigan, 311 West Hall, 1085 South University Ave., Ann Arbor, MI 48109-1107
   \and Eureka Scientific, 2452 Delmer St. Suite 100,Oakland, CA 94602, USA
   \and Observat\'orio do Valongo, Universidade Federal do Rio de Janeiro, Ladeira do Pedro Ant\^onio 43, Rio de Janeiro RJ 20080-090, Brazil
   \and Department of Physics \& Astronomy, University of California Los Angeles, 430 Portola Plaza, Los Angeles, CA, 90095-1547, USA
   \and Department of Astronomy, University of Washington, 3910 15th Avenue, NE, Seattle, WA, 98195, USA
   }

   \date{\today}

 
  \abstract
  {{Detection and study of the intracluster light in rich clusters of galaxies has been a problem of long standing challenge and interest.} Using the lowest surface brightness images of the Coma cluster of galaxies {in the $g$ and $r$ bands, from the Halos and Environment of Nearby Galaxies (HERON) Coma Cluster Project}, we obtained the most extensive image of intracluster light (ICL) in a single cluster to date, spreading over 1.5 Mpc from the cluster core. The unprecedented wealth of spectroscopic data {made publicly available by the Dark Energy Spectroscopic Instrument (DESI) Early Data Release, complemented with a compilation from the NASA/IPAC Extragalactic Database and the literature,} enabled the identification of 2,157 galaxy members within Coma, from which 42 distinct groups were identified. The synergy between these high-quality data allowed us to{: 1) calculate ICL fractions of $19.9\pm0.5$\% and $19.6\pm0.6$\% in the $g$ and $r$ bands, respectively, consistent with a dynamically active cluster, 2) unveil Coma's faintest tidal features, and 3) provide a comprehensive picture of the dynamics and interactions within this complex system}. Our findings indicate that the ICL connects several of these groups in a filamentous network, from which we infer the ongoing dynamical processes. In particular, we identified a faint stellar bridge linking the core of Coma with the galaxy NGC 4839, providing compelling evidence that this galaxy has already traversed the central region of the cluster. }


   \maketitle
%

\section{Introduction} \label{sect:intro}

The Coma cluster of galaxies (Abell 1656) is arguably one of the best studied structures in the Universe. It is a nearby ($z=0.023$) and massive cluster, with a halo mass of $M_{200}\approx 0.5-1.13 \times 10^{15}$ M$_{\odot}$ \citep{lokas2003,gavazzi2009,ho2022} . Its proximity and dimensions allow us to capture ongoing assembly processes in action and study them with an extraordinary level of detail. Recent works have shown that the current dynamical stage of Coma is turbulent: the currently accepted scenario describes two groups, associated with two elliptical galaxies located in the core of the cluster, NGC 4889 and NGC 4874, which are orbiting each other in the process of merging. The picture is completed with a third, further group, linked to the galaxy NGC 4839, which would be falling into the cluster potential \citep{briel1992,watt1992,davis1993,colless1996,arnaud2001,neumann2003,adami2005b,gerhard2007}. Multiwavelength evidence supporting this highly active scenario includes multiple components identified in the galaxy velocity distribution \citep{kent1982,fitchett1987,merritt1987,mellier1988,colless1996}; an elongated X-ray emission with substructures, which lacks significant temperature or metallicity gradients \citep{johnson1979,briel1992,edge1992,watt1992,davis1993,white1993,colless1996,vikhlinin1997,briel2001,arnaud2001,neumann2001,neumann2003,simionescu2013}; and the presence of a giant radio halo and a radio relic \citep{willson1970,jaffe1976,giovannni1985,kim1990,venturi1990,deiss1997,kronberg2007,brown2011}.\\

The most immediate consequence of these ongoing processes is that a significant amount of material is ejected from the interacting galaxies into the intracluster space. Low-surface-brightness features like the intracluster light (ICL), tidal streams, arcs, plumes, or loops (just to mention a few) are the remnants of galaxies' dynamical activity and, therefore, privileged witnesses of these interactions. The intracluster light, defined as the light emitted by stars that are gravitationally bound to the cluster potential but not hosted by any galaxy, is one of the best telltale signs of the past and ongoing assembly history of the host cluster, both at local and global scales {\citep{contini_review2021,contini2024chapter}}. These stars are stripped from their progenitor galaxies by a number of environmental-driven processes, such as major mergers with massive galaxies (including the BCG), cluster-cluster mergers, tidal stripping of luminous galaxies, preprocessing in infalling groups, and total disruption of dwarf galaxies {\citep[e.g., ][]{mihos2005,rudick2006,joo2023,demaio2018,ellien2019,jimenez-teja2021,montes_review2022,contini2024,jimenez-teja2024b,martis2024}}. Recent works consider that in situ star formation is also an important channel to produce ICL \citep{puchwein2010,sun2010,barfety2022}. These mechanisms act on different scales (cluster scale or locally) and with different degrees of intensity as a function of the clustercentric radius. The mechanisms that play the leading role depend on several factors, such as the dynamical stage of the host cluster or the redshift, and leave distinct imprints in the morphology, color, age, and metallicity of the ICL, and also in the ICL fraction \citep{jimenez-teja2018,jimenez-teja2021,deoliveira2022,dupke2022,jimenez-teja2023,jimenez-teja2024}. Therefore, a detailed knowledge on the ICL and its characteristics enables to read the past and present chapters of the assembly history of the cluster. {The study of these features is more accessible in nearby clusters, for which deeper information can be extracted from the ICL \citep{iodice2017,mihos2017,jimenez-teja2019,gu2020,kluge2021,arnaboldi2022,spavone2024}.} \\

The Halos and Environments of Nearby Galaxies (HERON) Survey \citep{rich2017,rich2019} is a project that uses a fully-dedicated 0.7-m telescope (see Sect. \ref{sect:data:imaging}) to obtain low surface brightness observations of more than a hundred {nearby galaxies mostly located in the Local Volume within 50 Mpc}, with the ultimate goal of reconciling theoretical predictions and observational measurements at the low surface brightness regime \citep{rich2012,muller2019}. The HERON Coma Cluster Project (described in a companion paper, Román et al., in prep) is a spin-off project that targets the Coma cluster of galaxies with the same instrument and a similar observational set up. As a result, it has obtained extremely deep images of the Coma cluster in the $g$ and $r$ bands (maximum limiting surface brightness of $\mu_g=30.1$ and $\mu_r=29.6$ mag arcsec$^{-2}$), unveiling a myriad of low-surface-brightness structures and details that had remained hidden so far. In this paper, we use these images of unprecedented depth to explore the dynamical history of galaxies and the most recent chapters of the assembly history of Coma, via the stellar material that is expelled during the numerous interactions that happen among the cluster members, i.e.,  the ICL. \\

This paper is organized as follows. Sect. \ref{sect:data} describes the imaging and spectral data used for this work. In Sect. \ref{sect:analysis}, we explain the analysis made with these data, including the process to generate the ICL maps and ICL fractions, the determination of the cluster membership, and the identification of galaxy groups in Coma. Sect. \ref{sect:discussion} discusses the obtained results to infer the dynamical state of the cluster and the recent dynamical history of its brightest galaxies and associated groups. We finally summarize our main conclusions in Sect. \ref{sect:conclusions}. Throughout this paper, we will assume a standard $\Lambda$CDM cosmology with $H_0=70$ km s$^{-1}$ Mpc$^{-1}$, $\Omega_m=0.3$, and $\Omega_{\Lambda}=0.7$. All the magnitudes are referred to the AB system.\\

\section{Data} \label{sect:data}

We used the images taken by the HERON Coma cluster project to calculate the ICL maps and the spectra observed by DESI to identify galaxy members. Although DESI provides the main bulk of the spectroscopic sample, we compiled additional spectra from the NED, the SDSS, and the literature (described in Sect. \ref{sect:analysis:cluster_membership}).\\

\subsection{Imaging: the HERON Coma cluster project} \label{sect:data:imaging}

{The Coma cluster (with center at R.A.: 12$^{\rm h}$59$^{\rm m}$44.40$^{\rm s}$, Dec.:+27$^{\circ}$54'44.9'' and redshift $z=0.0234$) is the focus of the HERON Coma Cluster project, an extensive observational campaign on the Coma cluster within the HERON survey \citep{rich2017,rich2019}. The HERON survey uses the Jeanne Rich telescope located at the Polaris Observatory Association in Frazier Park, California \citep{rich2012}. The telescope is a Centurion 28-inch hyperbolic primary mirror telescope with the CCD mounted at prime focus behind a 2-element Wynne corrector.  The 71-cm aperture with f/3.2 gives a field of view of 56.6 arcmin. The FLI ML09000 CCD camera has 12 $\mu \rm m$ pixels, giving a scale of 1.114 arcsec/pix.}\\


{Data processing was carried out with the objective of providing the highest reliability in the extreme low surface brightness regime. For this purpose, the images were first bias subtracted with standard methods. Flat-field correction was carried out by constructing a flat with strongly masked science images, which after further combination provided a high quality flat. The astrometric solution was obtained for all images by obtaining a first approximation using the Astrometry.net package \citep{lang2010}, being subsequently refined with the SCAMP software \citep{bertin2006}. The sky subtraction and coadding of the images was carried out by an iterative process, providing highly efficient final coadds at low surface brightness. This process consists of first obtaining a provisional coadd with a sky subtraction by removing only one pedestal value from each image. This provisional coadd, while containing all the low surface brightness information by not applying an aggressive sky background subtraction, contains multiple residuals of the light gradients from flat processing and atmospheric emission. We apply a very aggressive mask to this coadd by combining masks obtained with SExtractor and Noisechisel \citep{bertin1996,akhlaghi2015,akhlaghi2019}.  Subsequently, we applied this mask obtained from the coadd to each of the individual exposures, providing a much more efficient mask than if it had been applied to the coadd. Sky subtraction was performed using Zernike polynomials \citep{zernike1934}, which efficiently describe the wavefront deformations of an optical system. After sky subtraction of all individual exposures, we performed the photometric calibration using the Dark Energy Camera Legacy Survey \citep[DECaLS, ][]{dey2019} images as reference with a zero point of 22.5 mag. These sky-subtracted and photometrically calibrated individual exposures are combined by 3-sigma clipping averaging, also weighted by the signal-to-noise of each exposure, producing a coadd with a pixel scale of 1.114 arcsec. We perform several iterations using the coadd obtained in each step by repeating this process of coadding, masking, sky subtraction to individual exposures and combination. For this, in the first iterations we use low Zernike polynomial orders, until we reach a satisfactory solution in which the coadd does not present visually important gradients in the individual exposures. For the vast majority of the exposures, an order of 2 is sufficient, although in a few cases we use orders up to 4 for the Zernike polynomials, performing an individual visual inspection of each exposure to be sure that no strong gradients are present in the images. This process is performed for each of the \textit{g} and \textit{r} bands.}\\

{We also performed a subtraction of the brightest stars in the field that interfere or overlap with the regions of interest such as stellar and intracluster light halos in the Coma region. For this purpose, we perform a modeling of the point spread function of the telescope by observing very bright stars, using techniques similar to \cite{roman2020} and \cite{infante-sainz2020}. This subtraction of the brightest stars provides the possibility of studying the low surface brightness regions without the interference of scattered-light halos, very problematic in the Coma region.}\\

{The effective exposure times after having eliminated a considerable number of poor quality exposures, mainly due to the presence of the moon or very strong gradients in the images, is 38.4 and 48.25 h in the $g$ and $r$ bands respectively. The depth of the images provide a superb surface brightness limit in the region of maximum exposure of 30.1 and 29.8 mag arcsec$^{-2}$ \citep[$3\sigma$, measured in regions of $10''\times10''$, see ][]{roman2020} in the $g$ and $r$ bands, respectively. This depth is variable along the $1.5^{\circ}\times 1.5^{\circ}$ footprint centered on the Coma cluster, decreasing towards the outer parts of this region.}\\

{A comprehensive description of the HERON Coma Cluster Project explaining in detail the observations and data processing will be published soon by Román et al. (in prep). The paper by \cite{roman2023} provides a first result of the HERON Coma Cluster Project. This consists on the discovery of the Giant Coma Stream, being the stellar stream with the lowest surface brightness ($\mu_{g,max}=~29.5$ mag arcsec$^{-2}$) and largest size ($\sim 510$ kpc) discovered to date.}\\

\begin{figure*}
\includegraphics[width=\textwidth,valign=t]{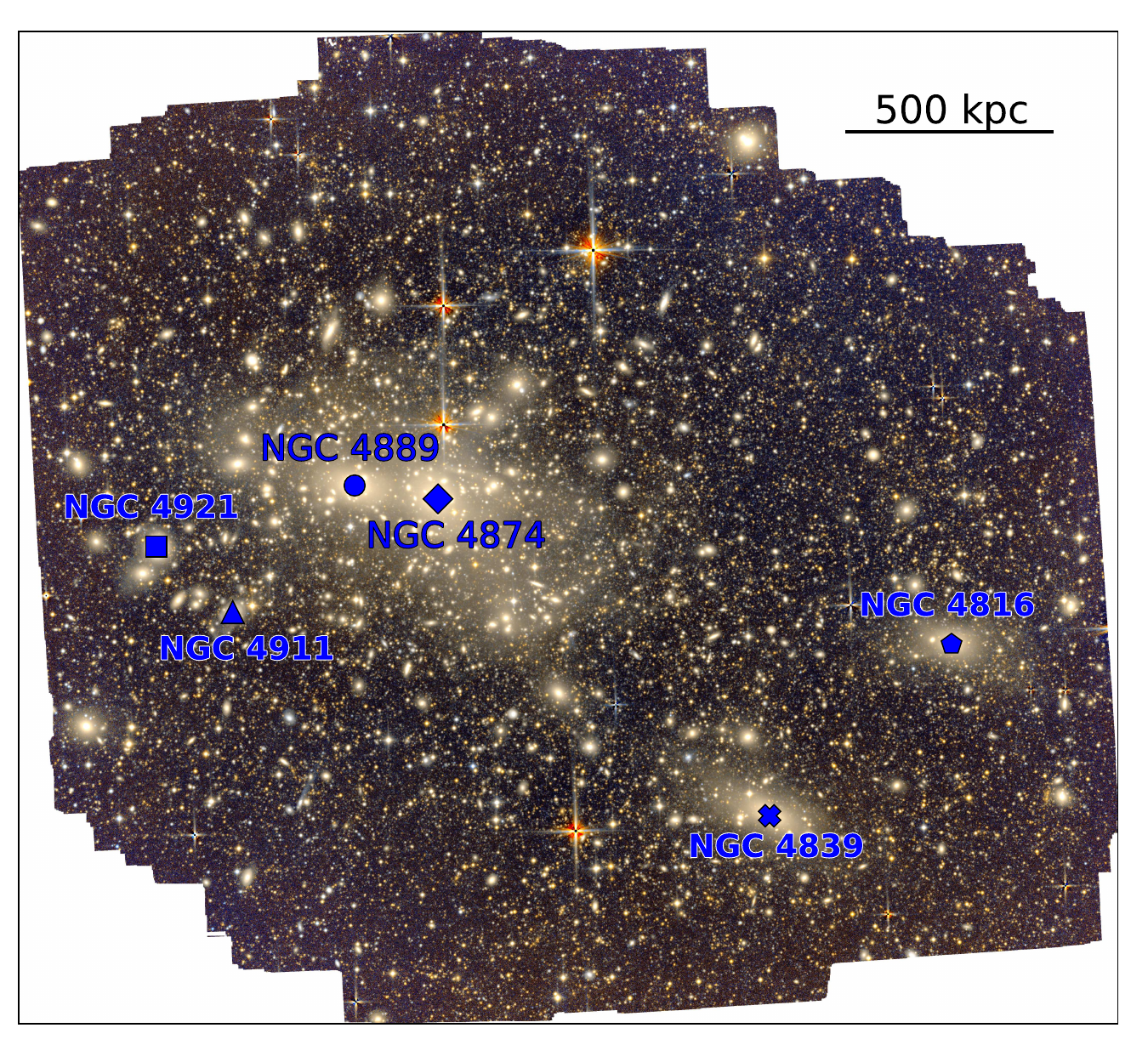} 
\caption{False color image of the HERON Coma field. Some of the brightest galaxies in the field are indicated in blue. At Coma's redshift, 500 kpc are equivalent to 18'' approximately. North is up, east is left.} 
\label{fig:color:original}
\end{figure*}

A false color image generated from the two broad bands can be observed in Fig. \ref{fig:color:original}. {This image is cropped to show a minimum of 1.5 hours of exposure time, corresponding to a surface brightness limit of at least 28.3 and 27.9 mag arcsec$^{-2}$ in \textit{g} and \textit{r} bands, respectively.}\\

\subsection{Spectra: DESI} \label{sect:data:spectra}

The Dark Energy Spectroscopic Instrument \citep[DESI, ][]{DESI2022}, mounted on the 4 m Mayall Telescope at Kitt Peak National Observatory, started operations in 2021. DESI is conducting an ongoing 5-year flux-limited survey that will obtain spectroscopic measurements of 40 million galaxies and quasars. In 2023, DESI made its first public release of ~1.8 million science spectra, the Early Data Release \citep[EDR, ][]{DESI2024}, collected during the five-month Survey Validation. The EDR includes observations done during the Target Validation Survey (SV1) and the One-Percent Survey (SV3). The One-Percent Survey (SV3) is a program conducted as the last phase of the Survey Validation to cover major datasets selected from other large past, ongoing, and future surveys. SV3 covers a total of $\sim 175$ square degrees split into 20 different locations in the sky. Two of these locations are the Coma cluster and its outskirts, which correspond to sets SV3 R4 and SV3 R16. Each location was observed a minimum of 12 times to {guarantee a lower limit of 95\% in the fiber assignment completeness for all targets and, specifically,} a 99\% completeness for the Bright Galaxy Survey (BGS). The BGS targets galaxies brighter than $r<19.5$ mag at $z<0.6$ along with a fainter, color-selected sample ($19.5<r<20.175$) and some low-redshift quasars \citep{hahn2023}. The BGS is designed to achieve a minimum redshift completeness of 98.5\% up to a magnitude of $m_r\geq 20.175$. In our analysis, we will assume this lower level of completeness for the whole set of spectra available for the Coma cluster and propagate the corresponding error to the final ICL fractions (see Sect. \ref{sect:analysis:ICLfractions}). \\

In total, DESI has collected and publicly released 167,059 spectra in an asymmetric region around Coma that reaches a maximum radius of $\sim 6.2$ deg ($\sim 10.36$ Mpc, at the redshift of the cluster) in the South-North direction and $\sim 3.2$ deg in the East-West direction. We will complement this sample later with a compilation of archival data (see Sect. \ref{sect:analysis:cluster_membership}), although for the subsequent analysis we will consider the completeness limits of the DESI observations.\\

\section{Analysis} \label{sect:analysis}

{In this section, we describe the steps followed to derived the ICL maps of Coma which, in combination with a detailed analysis and identification of the cluster member galaxies and their association in different subgroups, allowed us to understand the dynamics of the system. Finally, the flux contained in the different cluster components (ICL and galaxies) is used to calculate the ICL fraction, which provides additional information about the dynamical state of the cluster.}\\

\subsection{Calculation of the ICL maps} \label{sect:analysis:ICLmaps}

We generated the ICL maps of the Coma cluster with the CHEFs Intracluster Light Estimator \citep[CICLE, ][]{jimenez-teja2016}. CICLE removes the light from all the galaxies that lie in the field of view by fitting models with mathematical bases built with Chebyshev rational functions and Fourier series \citep[CHEFs, ][]{jimenez-teja2012}. Every galaxy is modeled up to the radius where its halo either converges asymptotically to a constant level or submerges into the background level. For the particular case of the brightest cluster galaxy (BCG), this limit is difficult to define because the extended halo usually associated with these massive galaxies diffuses progressively into the ICL. This limit can be estimated with the aid of a curvature map. Curvature maps measure the change in the slope of a surface at every point, so the transition between the BCG and the ICL can be identified with the curve of points where the curvature is maximum. Testing CICLE against simulations, we found that this transition was calculated with high precision, yielding a maximum error of 1\% in the ICL measured in clusters at $z\sim 0.25$. Further details about CICLE can be found in \cite{jimenez-teja2016}.\\

The accuracy and reliability of CICLE are also backed up by the result of a recent blind challenge carried out to compare different observational techniques to measure the ICL, using mock images with the observational characteristics expected for the Vera C. Rubin Observatory's Legacy Survey of Space and Time after ten years of operations \citep{brough2024}. Using projected images of clusters simulated using four different suites of cosmological hydrodynamical simulations, observer's measurements of the (BCG+)ICL fraction (defined as the ratio of (BCG)+ICL to the total luminosity of the cluster) were compared to those of the simulators, who performed their calculations in the original three dimensional space of particles. Among all observational methods tested, CICLE yielded the best results for both fractions and the smallest scatter, probing simultaneously its accuracy and robustness. Additionally, CICLE was among the techniques that suffered less impact from projection effects, certifying its reliability.\\

It is important to note that, while CICLE fits and removes the galactic light to leave only the ICL and background, the detection of the sources (stars and galaxies) must be carried out previously with a different software, to feed their coordinates to CICLE. We applied SExtractor \citep{bertin2010} for this matter. The exceptional depth of the HERON images of Coma made it very complex to find an optimal configuration for SExtractor, so several runs of SExtractor+CICLE were necessary to completely remove all the galactic light. We first detected and fitted the brightest and most extended galaxies (BACK\_FILTERSIZE=3, BACK\_SIZE=128, DEBLEND\_MINCONT=0.1, DEBLEND\_NTHRESH=32), while more compact and less luminous sources were modeled in a second step (BACK\_FILTERSIZE=3, BACK\_SIZE=25, DEBLEND\_MINCONT=0.01, DEBLEND\_NTHRESH=64). The resulting image was a combination of ICL and background. To remove the latter, we estimated it with NoiseChisel \citep{akhlaghi2015,akhlaghi2019} using the {default} configuration. \\

\begin{figure*}
\includegraphics[width=\textwidth,valign=t]{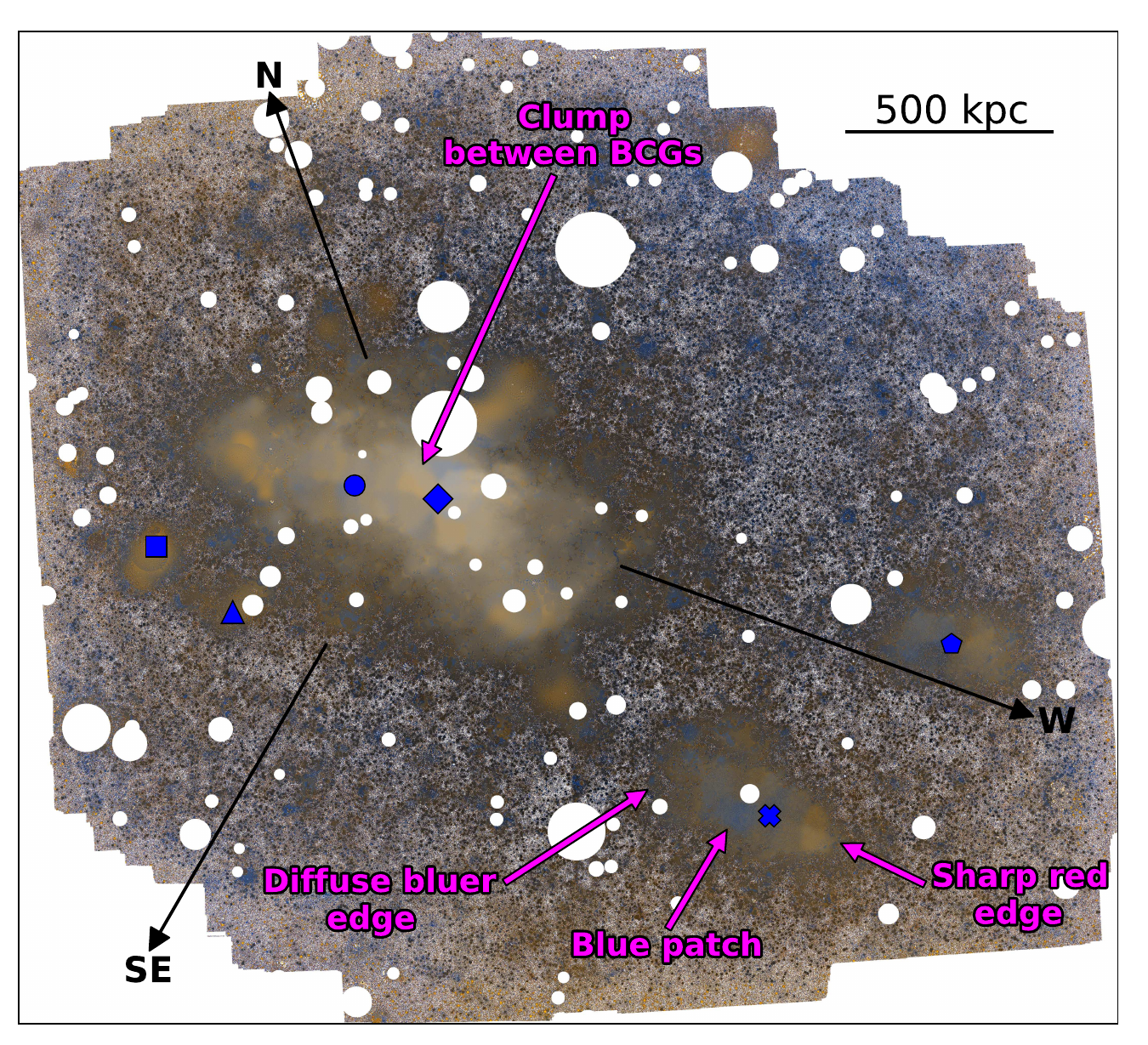}
\caption{False color image of the ICL in the region showed in Fig. \ref{fig:color:original}. The background is the composite $\overline{g+r}~$  ICL map in grayscale, to facilitate the visualization of the lowest surface brightness regions of ICL. White circles are masked stars. The blue symbols correspond to the brightest galaxies labeled in Fig. \ref{fig:color:Coma}. Black arrows show the direction of the intracluster filaments identified by \cite{hyeonghan2024} through weak lensing and named by them as N, W, and SE filaments. Magenta labels indicate ICL features that are described in Sect. \ref{sect:discussion:BGCs}, \ref{sect:discussion:infalling_groups}, and \ref{sect:discussion:NGC4839}. North is up, east is left.} 
\label{fig:color:ICL}
\end{figure*}

A color composite image of the resulting ICL maps, made from the $g$ and $r$-bands, can be observed in Fig. \ref{fig:color:ICL}. We observe by eye the ICL spreading out up to an extraordinary radius of $\sim 1.5$ Mpc, along with several substructures, many of them already identified in previous works \citep[e.g., ][]{adami2005a,adami2005b,jimenez-teja2019}. To the best of our knowledge, this is the largest radius up to which the ICL has been detected in an individual cluster, only surpassed by the analysis of \cite{gonzalez2021} on MACS J1149.5+2223, who were able to reconstruct the BCG+ICL surface brightness profile out to a radius of $\sim 2$ Mpc. Unfortunately, an image of the ICL is not shown in \cite{gonzalez2021}; instead, the detection is derived from a median-averaged profile in circular bins.\\


\subsection{Cluster membership} \label{sect:analysis:cluster_membership}

In order to identify cluster members, we used the redshift catalogs produced by the DESI collaboration, in particular, the zall-pix-edr-vac catalog\footnote{https://data.desi.lbl.gov/public/edr/vac/edr/zcat/fuji/v1.0/}. This catalog achieves full-depth by combining all exposures for targets on a given HEALPix \citep{gorski2005} pixel of the sky \citep{DESI2024}. Running a search of 6.2 deg around the center of the Coma cluster (chosen as the mean point between the two BCG peaks, i.e., R.A.: 12$^{\rm h}$59$^{\rm m}$51.61$^{\rm s}$, Dec.:+27$^{\circ}$57'59.18''), we found 167,059 spectroscopic redshifts. However, these redshifts include fibers pointing to the sky as well as galaxies targeted several times in different locations. To purge the catalog from these duplications and sky targets, we downloaded the photometric catalogues published as part of the Legacy Survey Data Release 10\footnote{https://www.legacysurvey.org/dr10/description/} and matched them to the spectroscopic catalog of DESI. After this step, we ended up with 118,967 galaxies with spectroscopic redshifts in the Coma cluster field and outskirts. We then proceeded to filter the catalog to remove all objects with an insecure redshift estimation, remaining 103,179 reliable spectra. \\

We used the minimum and maximum coordinates of the region covered by DESI around Coma to define a rectangular search region, which we later fed into the NASA/IPAC Extragalactic Database (NED) and the Sloan Digital Sky Survey Data Release 13 \citep[SDSS-IV DR13, ][]{blanton2017,albareti2017} to complement the DESI sample. We found 985 (NED) and 13 (SDSS) additional galaxies with secure spectroscopic redshift. Additionally, \cite{healy2021} compiled spectroscopic redshifts from 30 works in the literature along with their own observations in a 2 deg radius around Coma, from which  we added 9 objects to our previous sample. Finally, we ended up with a total of 104,177 spectroscopic redshifts in a region of $6.2^{\circ}\times 3.2^{\circ}$.\\


We then applied a rough cut of $\pm 5500$ km/s around the mean velocity of the Coma cluster, yielding 3,104 galaxies. We refined this selection using the caustic technique \citep{kaiser1987,regos1989,diaferio1997,diaferio1999,serra2013}, implemented in the publicly available Python package CausticSNUpy \citep{kang2024}. When galaxy members of a cluster are plotted in a phase-space diagram (i.e., the plane of the line-of-sight velocity of the galaxies in the cluster rest frame versus their projected distance from the cluster center), they distribute in trumpet-like shape, whose boundaries depend on the escape velocity profile of the cluster and the velocity anisotropy profile. These boundaries are called caustics and all galaxies that lie within the region delimited by them are identified as members while those outside are classified as outliers. The caustic technique is proved to be effective up to $3r_{200}$. \\

\begin{figure}
\centering
\includegraphics[width=0.48\textwidth]{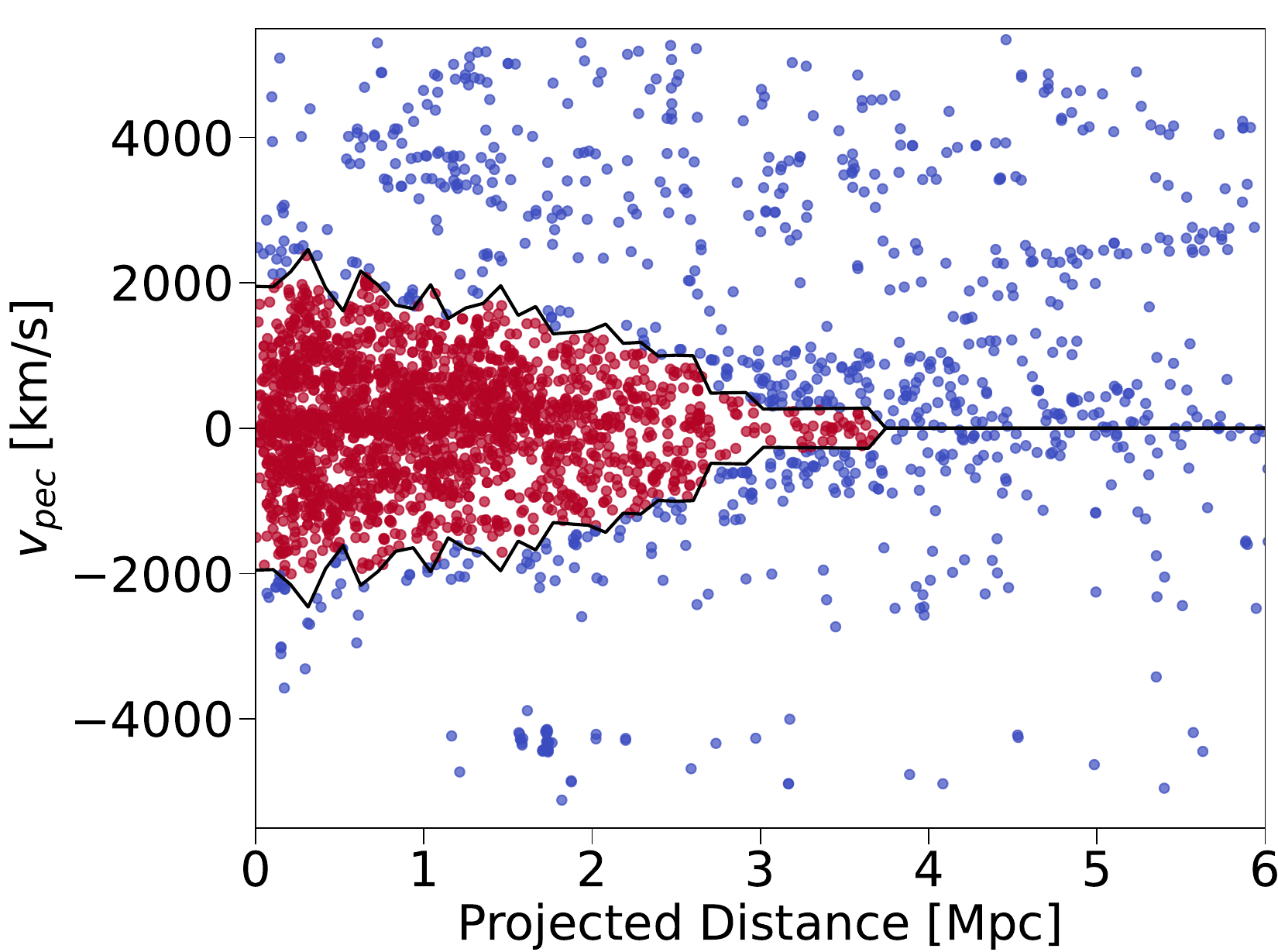}
\caption{Phase-space diagram for all galaxies in a rectangular region of $6.2^{\circ}\times 3.2^{\circ}$ ($\sim 10.36\times 5.35$ Mpc$^2$) around the center of Coma and in the redshift range $|z-0.023|\leq 0.018$. We have cropped the plot at 6 Mpc to facilitate visualization. Galaxy members identified by the caustic technique, are plotted in red, while outliers are colored in blue.} 
\label{fig:caustic}
\end{figure}

\begin{figure*}
\includegraphics[width=\textwidth,valign=t]{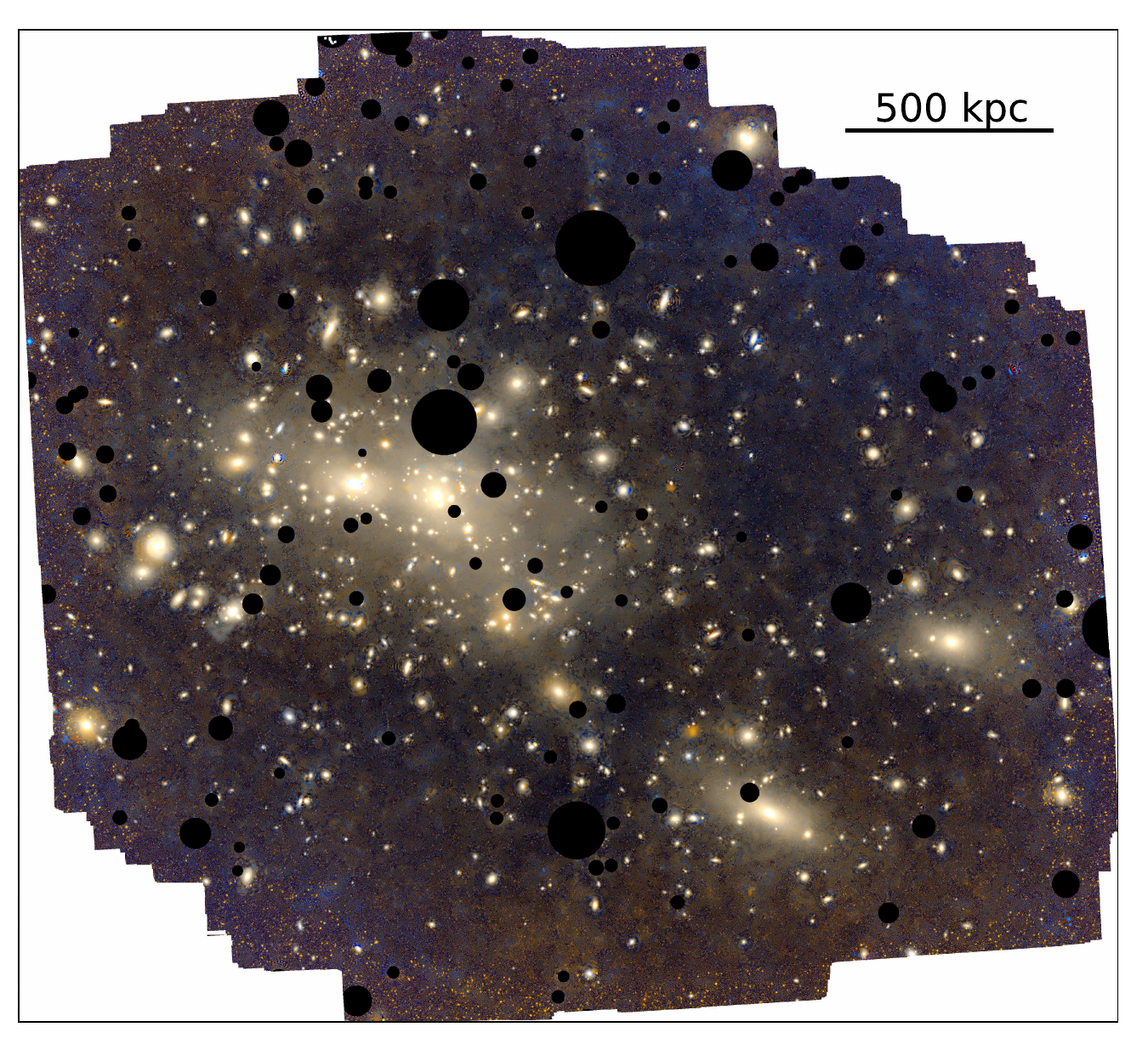}\\
\caption{False color image of the Coma cluster of galaxies. The image shows the ICL and the galaxies spectroscopically classified as members in the region showed in Fig. \ref{fig:color:original}. Black circles are masked stars. North is up, east is left.}
\label{fig:color:Coma}
\end{figure*}

The phase-space diagram of Coma can be observed in Fig. \ref{fig:caustic}, {computed using (R.A., Dec.)=$(194^{\circ}95305, 27^{\circ}98069)$ and $v^{cluster}=6951$ km/s, as the center and mean velocity of the cluster, respectively}. We identified 2,157 galaxy members within a radius of $\sim 3.8$ Mpc, being $\sim 39$\% of them located in the field of view of our HERON images {(see catalog in Appendix \ref{appendix:catalog})}. We estimated a velocity dispersion of member galaxies of $795\pm 11$ km/s. Since the caustic is related to the escape velocity from the gravitational potential well generated by the cluster, the cluster mass can be inferred from it under the assumption of spherical symmetry. We estimated a cluster mass of $1.22\times 10^{15}$ M$_{\odot}$ for Coma, in nice agreement with the dynamical mass of $1.26^{0.52}_{-0.37}\times 10^{15}$ derived by \cite{ho2022} using deep learning. A false color image of the Coma cluster members that lie within the field of view of the HERON observations is displayed in Fig. \ref{fig:color:Coma}.\\

\subsection{Identification of substructures} \label{sect:analysis:DS+test}

Identifying (infalling) groups in clusters is essential to get a complete picture of the processes that govern the ICL formation, because these are possible sites of galaxy pre-processing and in-turns of intergalactic material and tidal dwarf galaxies, all channels of ICL injection. This is clearly exemplified by the neighboring cluster Abell 1367 \citep{gavazzi2003}. \\

The unprecedented wealth of spectroscopic data in the region of Coma and its outskirts provided by DESI enabled to search for possible substructures in the distribution of galaxy members. Among all publicly available techniques in the literature, we chose the DS+ \citep{benavides2023}, a new version of the traditional method of \cite{dressler1988} that gathers in a single code all different improvements and extensions made to the original algorithm in the last years. DS+ not only identifies substructures in clusters, but also characterizes them by assigning a probability to each individual galaxy of belonging to a certain group. The DS+ method compares the local velocity field of galaxy members, estimated from multiple numbers of neighbors, with the mean velocity and velocity dispersion radial profile of the whole cluster. Significant differences between local and the global velocities would be suggestive of the presence of a group. Applying a certain number of Monte Carlo resamplings, the DS+ method assigns a probability to the local velocity dispersions, which is translated into a probability for each individual galaxy of belonging to a certain group. Under the no-overlapping mode, groups are uniquely characterized, that is, a single galaxy cannot be assigned to two different substructures, and those groups that are close spatially and with similar velocities are merged to minimize fragmentation. \\

\begin{table}
\caption{Properties of the 42 groups identified by the DS+ method and plotted in Fig. \ref{fig:DS+}}
\centering
\begin{tabular}{cccccc}
ID & Ngal & Dist &  $\sigma$ &  $v_{mean}$ & $v_{mean}-v^{cluster}$\\
 & & (kpc) & (km/s) & (km/s) & (km/s)\\
 \hline
1 & 6 & 4841.63 & 117.46 & 6771.31 & -211.78\\
2 & 30 & 2404.11 & 638.75 & 7327.62 & 344.53 \\
3 & 15 & 1808.61 & 456.74 & 7508.20 & 525.11\\
4 & 30 & 3255.26 & 557.59 & 6608.02 & -375.07\\
5 & 24 & 2148.00 & 788.24 & 6497.25 & -485.84\\
6 & 30 & 1548.39 & 488.86 & 6852.65 & -130.44\\
7 & 24 & 2753.19 & 370.18 & 7187.89 & 204.8\\
8 & 30 & 835.61 & 612.04 & 6526.38 & -456.71\\
9 & 12 & 1414.54 & 883.50 & 7647.70 & 664.61\\
10 & 6 & 1491.67 & 168.49 & 6250.67 & -732.42\\
11 & 12 & 1735.66 & 375.05 & 7625.72 & 642.63\\
12 & 27 & 2924.02 & 404.07 & 6885.34 & -97.75\\
13 & 6 & 4836.71 & 114.31 & 6950.68 & -32.41\\
14 & 18 & 1195.18 & 480.50 & 6646.06 & -337.03\\
15 & 6 & 4903.15 & 136.41 & 6933.19 & -49.9\\
16 & 9 & 688.56 & 369.53 & 7394.21 & 411.12\\
17 & 21 & 1333.22 & 494.02 & 7341.91 & 358.82\\
18 & 21 & 120.46 & 592.17 & 6672.66 & -310.43\\
19 & 6 & 2017.39 & 710.56 & 7998.46 & 1015.37\\
20 & 9 & 401.00 & 522.03 & 7778.28 & 795.19\\
21 & 21 & 351.61 & 847.57 & 7633.28 & 650.19\\
22 & 6 & 491.07 & 223.00 & 8051.42 & 1068.33\\
23 & 6 & 652.71 & 434.89 & 5725.03 & -1258.06\\
24 & 27 & 88.57 & 628.95 & 7183.91 & 200.82\\
25 & 12 & 2107.20 & 309.17 & 7204.51 & 221.42\\
26 & 9 & 1515.95 & 914.41 & 6184.71 & -798.38\\
27 & 6 & 3032.65 & 304.53 & 6182.71 & -800.38\\
28 & 9 & 3298.37 & 593.80 & 6306.96 & -676.13\\
29 & 9 & 1117.26 & 220.07 & 6775.97 & -207.12\\
30 & 9 & 1211.88 & 253.06 & 7350.91 & 367.82\\
31 & 12 & 399.13 & 398.26 & 6549.21 & -433.88\\
32 & 6 & 216.08 & 162.23 & 7330.42 & 347.33\\
33 & 6 & 462.28 & 367.68 & 5932.39 & -1050.7\\
34 & 6 & 811.63 & 207.87 & 7466.83 & 483.74\\
35 & 18 & 396.26 & 543.82 & 6613.58 & -369.51\\
36 & 15 & 1449.81 & 434.04 & 7276.76 & 293.67\\
37 & 9 & 264.97 & 374.45 & 7181.02 & 197.93\\
38 & 9 & 2871.31 & 729.52 & 6407.56 & -575.53\\
39 & 6 & 109.17 & 276.12 & 7141.05 & 157.96\\
40 & 12 & 257.56 & 960.23 & 6206.70 & -776.39\\
41 & 12 & 521.82 & 938.69 & 6149.49 & -833.6\\
42 & 18 & 1484.29 & 423.29 & 6950.52 & -32.57\\
\end{tabular} 
\tablefoot{Columns correspond to: group identifier, number of members, distance to the center of the cluster, velocity dispersion, mean velocity, and relative mean velocity with respect to that of the cluster.}\label{table:DS+}
\end{table}

\begin{figure*}
\centering
\includegraphics[width=\textwidth]{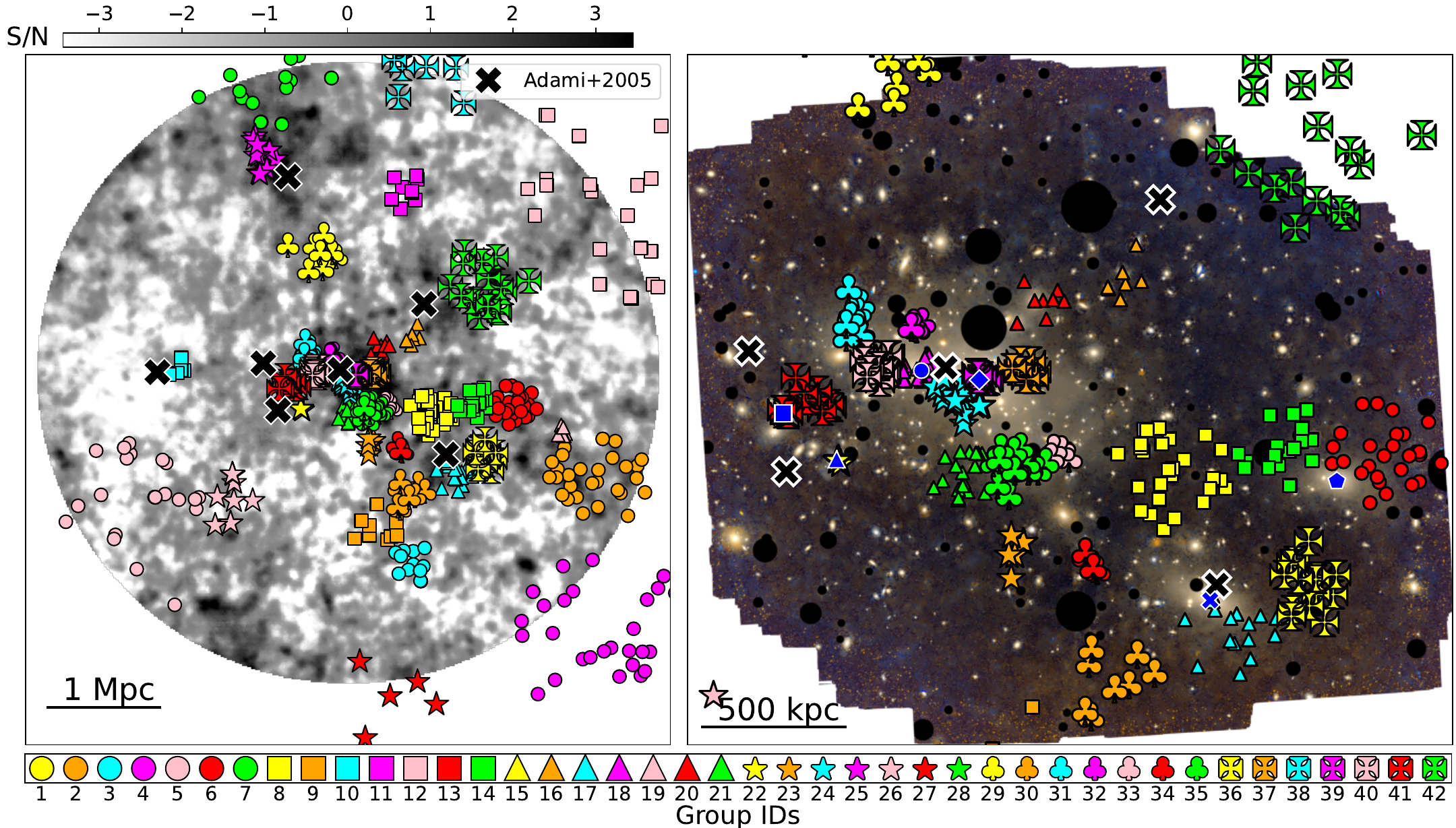}
\caption{Spatial distribution of the groups identified in Coma. Left: weak lensing mass of Coma \citep[courtesy of ][see their Fig. 1]{hyeonghan2024}, with the projected spatial distribution of the 42 groups identified by the DS+ method. Black crosses with a white edge indicate the location of the groups with more than four members identified by \cite{adami2005b}. Right: groups identified by DS+ within the field of view of our HERON images of Coma, plotted over a pseudo-color image of the Coma members. The blue symbols with a white edge indicate the location of the brightest galaxies that were labelled in Fig. \ref{fig:color:original}, using the same markers. North is up, east is left.} 
\label{fig:DS+}
\end{figure*}

Following the directions of \cite{benavides2023}, we ran DS+ over the catalog of 2,157 Coma members previously identified, with 1000 Monte Carlo resamplings and a maximum probability of 0.5, a similar set-up of previous works in the literature \citep[e.g., ][]{kim2024}. We discarded all groups with less than five members to keep just those with the highest significance, yielding a total of 42 groups identified within a radius of 3.8 Mpc whose properties are listed in Table \ref{table:DS+}. {These properties are directly derived by DS+ using the gapper method \citep{benavides2023,beers1990,girardi1993} and are: the number of galaxies that belong to the group, the distance of the group to the center of the cluster, the velocity dispersion, the mean velocity, and the difference between the mean velocity of the group and that of the cluster}. The projected spatial distribution of the groups and their galaxy members can be observed in Fig. \ref{fig:DS+}, plotted over the weak lensing mass map computed by \cite{hyeonghan2024} (left) and over a false color image of the Coma cluster members (right). We also indicate with yellow crosses the approximate location of the groups with at least five members previously identified by \cite{adami2005b}, using the \cite{serna1996} hierarchical method. We find a good agreement between the common substructures identified by both methods, observing that all groups identified by \cite{adami2005b} have a close group identified by DS+. More recently, \cite{healy2021} applied the classical \cite{dressler1988} algorithm to a catalog of 1,095 spectroscopically confirmed galaxy members within a radius of $\sim 3.4$ Mpc around Coma center, finding 15 groups with more than 5 objects. A detailed close look to Fig. 11 in \cite{healy2021} confirms that we also identified these 15 substructures from our data. \\

\subsection{ICL fraction and error} \label{sect:analysis:ICLfractions}

The ICL fraction is defined as the ratio of the ICL to the total cluster (ICL plus galaxies) flux. As opposed to aperture photometry, we estimate our ICL fractions up to the maximum radius where the ICL is detected. The error of the ICL fraction comes from the combination of three different sources: 1) the photometric error (standard error associated to any flux measurement), 2) the geometrical error (error made by CICLE in the determination of the transition from the BCG- to the ICL-dominated region along with the missing flux from the outskirts of the BCG halo that extend beyond this boundary, submerged into the ICL), and 3) the incompleteness in the cluster membership (error due to both fiber incompleteness during the acquisition of the spectra and faint galaxies that were not targeted by DESI). \\

The geometric error is calculated through simulations that mimic the BCG+ICL system of Coma, that is, with the same profiles, effective radii, and magnitudes. These simulations are polluted with Gaussian noise with a S/N comparable with that of the real images, and we build 100 different realizations of these noisy images to later run CICLE on them. The final error is the mean difference between the ICL fractions calculated by CICLE over the 100 simulations and the original ICL fraction computed directly from the clean BCG+ICL system. This process is repeated for each one of the two BCGs of Coma.\\

To estimate the error due to spectra incompleteness, we used the most conservative numbers provided by the DESI collaboration for the EDR, that is, that the DESI spectra is 98.5\% complete up to a magnitude $m_r\geq 20.175$ (see Sect. \ref{sect:data:spectra}). We note here, though, that this number may be overestimated for the particular case of Coma, given that the larger crowding in dense environments with respect to the field can lead to the physical incapability of the positioners to observe all targets \citep{lasker2024}. For example, the problem of fiber collisions affects targets two close in the sky and also the fact that only one DESI fiber can be placed within a single fiber patrol radius, having a small overlapping region with neighboring fibers. Additional problems that effectively decrease the spectroscopic completeness are power and software-based failures that may interrupt the fibers positioning, sky subtraction imprecision, atmospheric variability, and redshift determination errors. We alleviated some of these caveats by making the spectral coverage denser via including additional gathered from the NED and SDSS databases (see Sect. \ref{sect:analysis:cluster_membership}), although we cannot estimate how much we contributed to achieve the 98.5\% completeness in the region of Coma. Assuming that the 1.5\% of unobserved targets has an average flux comparable to the median of observed targets {and dividing their total flux by the sum of the flux of all identified cluster members}, we estimated that the flux missed is 0.37 and 0.34\% of the total cluster flux, in the $g$ and $r$ bands respectively.\\

Additionally, we estimated the impact that the flux-limited selection of the DESI spectra has on our measurement of the total luminosity of Coma, up to the limiting magnitudes of the HERON images. At the redshift of the Coma cluster, $z= 0.023$, the magnitude depth of the DESI spectra, $m_r=20.175$ translates into an absolute magnitude of $M_r=-14.84$ mag. We used the limiting surface brightness maps elaborated by \cite{roman2023} to calculate the average surface brightness limits within the region of detection of the ICL, yielding $\mu_{lim,g} = 29.7$ and $\mu_{lim,r}=29.3$ mag arcsec$^{-2}$ in the $g$ and $r$ bands, respectively. From these values, we derived limiting magnitudes of $m_{lim,g}=24.7$ and $m_{lim,r}=24.3$ mag in the $g$ and $r$ bands, respectively. These limits correspond to absolute magnitudes of $M_{lim,g}=-10.3$ and $M_{lim,r}=-10.7$ mag, which implies that the dwarf galaxies population is well mapped in our deep images. Using the luminosity functions of \cite{beijersbergen2002} for the Coma cluster, we inferred that  the total cluster flux is underestimated by $\sim 1$\% due to the limited depth of the spectra in both filters. \\

All these three sources of error were propagated to the final errors of the ICL fractions {in the standard way, that is, by calculating the partial derivatives with respect to each one of the variables. The final errors are} listed in Table \ref{table:data} along with the nominal values of the fractions {and the total magnitude of the ICL}. Our ICL fractions ($\sim 20$\% for both filters) are in the upper limit (or even higher) than those measured in previous works \citep[e.g., ][]{melnick1977,adami2005a,jimenez-teja2019}, evidencing the higher depth of our images. The errors are also smaller than those obtained in previous works, which is due not only to the high quality of the HERON Coma images but also to the deeper spectral coverage provided by DESI.\\


\begin{table*}
\caption{Depth of the HERON Coma images, and output parameters yielded by CICLE.}
\centering
\begin{tabular}{ccccc}
Filter & Surface brightness limit  & {ICL magnitude} & ICL fraction & Radius\\ 
   & (mag arcsec$^{-2}$) & {(mag)} & (\%) & (Mpc)\\
\hline
$g$ & ${30.1\pm 0.1}$ & ${10.64\pm 0.01}$ & ${19.9\pm 0.5}$ & 1.33\\ 
$r$ & ${29.6\pm 0.1}$ & ${9.85\pm 0.01}$ & ${19.6\pm 0.6}$ & 1.59\\
\hline
\end{tabular} 
\tablefoot{{For each band, we report the $3\sigma$-surface brightness limits calculated in boxes of $10\times10$ arcsec$^2$ by \cite{roman2020}, the ICL magnitude and fraction measured and the maximum radius of detection of ICL.}}\label{table:data}
\end{table*}

\section{Discussion} \label{sect:discussion}

\subsection{Dynamical state of Coma} \label{sect:discussion:dynamical_state}
\begin{figure}
\centering
\includegraphics[width=0.5\textwidth]{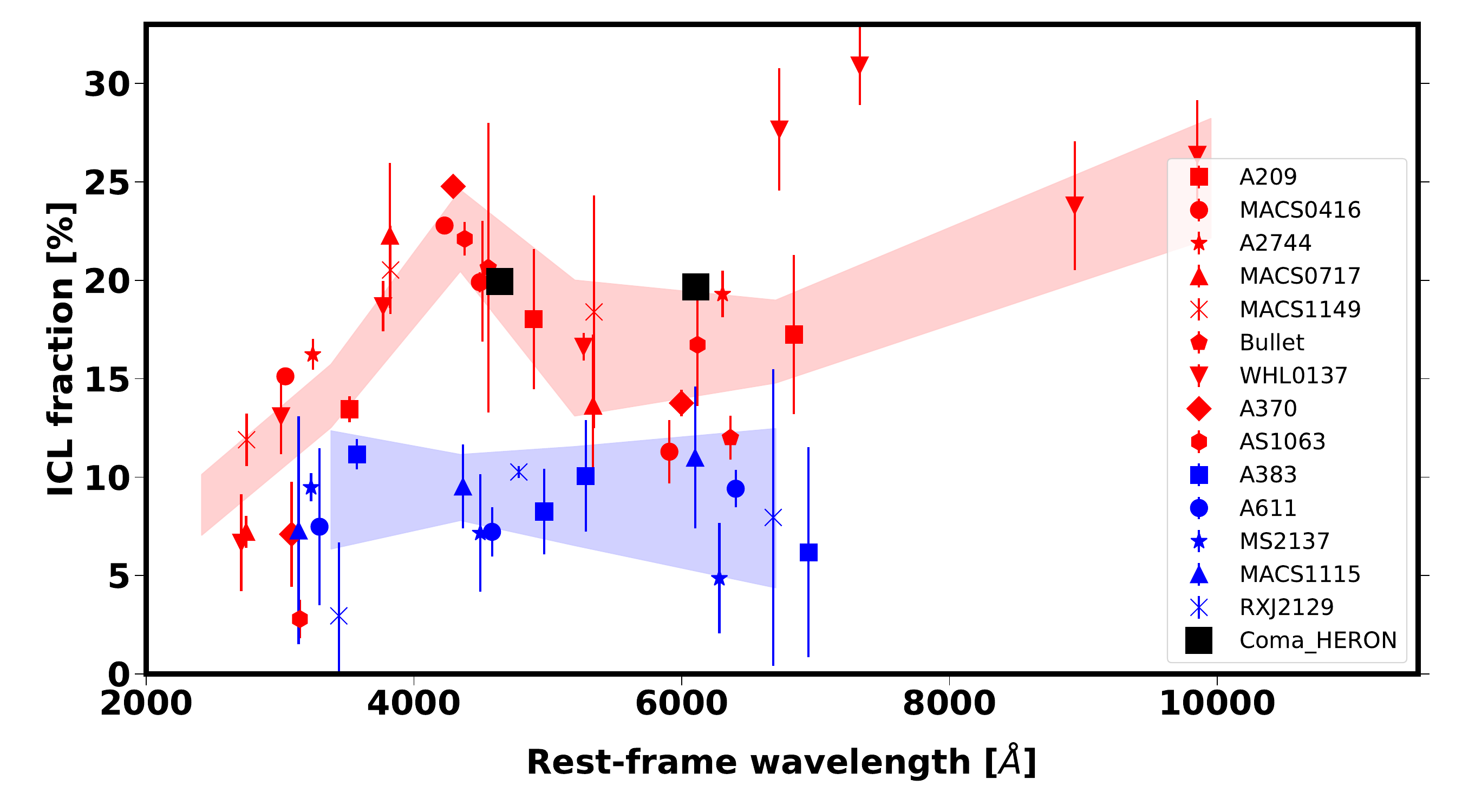}
\caption{ICL fractions measured for two samples of clusters at different wavelengths: relaxed (blue) and merging (red). Coloured shadowed regions indicate the error-weighted average of the ICL fractions measured for each one of the samples. Black squares are the ICL fractions measured for Coma in this work.} 
\label{fig:ICLfractions}
\end{figure}

Although Coma was considered an archetypal relaxed cluster for decades, it is well-known now that it is far from virialization and many substructures are in an ongoing process of merging \citep[see, e.g., ][ and references therein]{jimenez-teja2019}. The ICL fraction has been proven a highly reliable indicator of the dynamical state of clusters, both at low and high redshift \citep{jimenez-teja2018,jimenez-teja2021,deoliveira2022,dupke2022,jimenez-teja2023}. In Fig. \ref{fig:ICLfractions}, we plot the ICL fractions found for Coma with the HERON images along with ICL fractions measured by previous works for a sample of relaxed (blue) and merging (red) clusters. As we can see, they nicely lie within the expected region of merging clusters. As opposed to relaxed clusters that have nearly constant and low ICL fractions in the optical range, independently of the wavelength, the distribution of the ICL fraction in disturbed clusters is characterized by higher ICL fractions that peak between $\sim 3800-4800$ \AA~ (named IE in the literature, from ICL fraction excess) originated from an excess of younger and/or lower-metallicity stars (typically, A- and F-type stars) from the outskirts of galaxies, that are rapidly thrown into the ICL during the merger. The new ICL fractions thus ratify that Coma is in an unrelaxed state.\\

\begin{table}
\caption{Mean velocity of the Coma cluster and its brightest galaxies}.
\centering
\begin{tabular}{lc}
Object & $v$\\ 
 & (km/s)\\
\hline
Coma & 6983.1 \\
NGC 4874 & 7168.0\\
NGC 4889 & 6445.5\\
NGC 4839 & 7338.9 \\
NGC 4921 & 5471.2 \\
NGC 4911 & 7971.5 \\
\hline
\end{tabular} 
\label{table:mean_vel}
\end{table}

\subsection{Structure of the cluster core: the two BCGs} \label{sect:discussion:BGCs}

As previously noticed by several authors \citep[e.g., ][]{adami2005a,jimenez-teja2019}, the ICL in the core of Coma is heterogeneous and clumpy, with clumps of varied colors and brightness spread all around the two brightest galaxies, NGC 4874 and NGC 4889 (see Fig. \ref{fig:color:ICL}; these galaxies are marked with a blue circle and diamond, respectively). The main cloud of ICL is an elongated structure, the main axis of which, is denser and has an arc-like shape. The brightest concentration of ICL is the one that surrounds NGC 4874. The dynamics of the two main ellipticals in Coma has been intensively discussed in the literature \citep[e.g., ][ and references therein]{adami2005b,gerhard2007,gu2020}. The general consensus is that the two galaxies have small groups dynamically associated, which are in an ongoing merger that is very likely in a stage posterior to one or several core passages. We indeed find that groups \#39 and \#18 are associated with NGC 4874 and NGC 4889 respectively, with mean velocities consistent with those of these two elliptical galaxies (see Tables \ref{table:DS+} and \ref{table:mean_vel}). The mean velocity of group \#39 is closer to that of the cluster than the mean velocity of group \#18, suggesting that NGC 4874 has been in Coma longer and NCG 4889 is an infalling  galaxy, as hypothesized by \cite{adami2005b}. The fact that the highest amount of ICL is located around NGC 4874 is in agreement with this scenario, because the ICL is bound to the gravitational potential of the cluster and, therefore, concentrates around the potential well. Independently of which one is the colliding galaxy, it seems clear that both of them are in an advanced stage of merging, with dense clumps of ICL surrounding them. It is specially notable the ICL clump located right in between the two BCGs (see Fig. \ref{fig:color:ICL}), probably fed by stars that are freed from the galaxies' outskirts and the disruption of dwarfs due to the strong gravitational forces that dominate this region.\\ 

An independent group is identified in this region, the group \#24 (see Fig. \ref{fig:DS+} right). However, its mean velocity is very similar to that of the NGC 4874 group (\#39) and not so close to that of the cluster, which points to two plausible explanations: 1) either these two groups constitute a single structure and DS+ has incorrectly fragmented it, or 2) this group is in the process of being detached from its original, progenitor group \#39 (or, contrarily, group \#39 is being detached from \#24, given that its velocity dispersion is higher {by $\sim 350$ km/s, see Table \ref{table:DS+}}). A similar case is observed with group \#37, northwest to NGC 4874. {Its mean velocity is consistent with that of group \#39 (7181 and 7141 km/s, respectively)}, but it is still identified as an independent structure. With these two ``satellite'' groups located at both sides of group \#39 and embedded into a common cloud of ICL, the hypothesis that they could have originally been part of a larger structure (with at least 39 members), located at the bottom of Coma gravitational potential well and later displaced by the interaction with the NCG 4889 group, is reinforced. We may be witnessing the shredding and dissolution of the group caught in the act. Interestingly, group \#24 appears elongated in the direction of NGC 4839 (marked with a blue cross in Fig. \ref{fig:DS+}), which would be explained if this galaxy had already traversed the core of Coma in a relatively recent past (see Sect. \ref{sect:discussion:NGC4839}) altering the positions of the group \#24 galaxies. Indeed, this would also explain why group \#24 has a velocity consistent with that of NGC 4874. On the other hand, \cite{gerhard2007} found that the intracluster planetary nebulae (ICPNe) located in the southwestern end of the group have velocities consistent with that of the NGC 4889 structure. Probably, the two BCGs are spiraling around each other after second core passage and NGC 4889 left behind a substantial amount of ICL (including ICPNe) in its trajectory from the southwest \citep[see Fig. 2 in ][]{gerhard2007}. However, the crossing of the NGC 4839 group could have provoked the elongation of the \#24 group (and, possibly, its detachment from group \#39), along with a phase-mixing of the ICL \citep[indeed, the velocity distribution of the ICPNe located in this region is skewed and shows a secondary, bluer peak; see Fig. 1 left in ][]{gerhard2007}.\\

East to NCG 4889, DS+ identifies the group \#40, whose mean velocity suggests that is independent of NGC 4889 {(6206 km/s for the group and 6445 km/s for NGC 4889, see Tables \ref{table:DS+} and \ref{table:mean_vel}}. Its velocity dispersion is the highest of all groups detected, which indicates that it could be dissolving. Indeed, {Figs. \ref{fig:color:ICL} and \ref{fig:ICL+Xray} show} that the ICL has a lower surface brightness in this region (compared to the rest of the main cloud of ICL), which is consistent with the fact that group \#40 does not have much intragroup light (i.e., pre-processed ICL), given that high velocity dispersion does not favor galactic encounters and interactions.\\

\subsection{Structure of the cluster core: infalling groups} \label{sect:discussion:infalling_groups}

Fig. \ref{fig:ICL+Xray} shows the composite $\overline{g+r}~$ ICL map after applying a threshold segmentation to enhance the lowest surface brightness substructure. We plotted the ICL isocontours (blue), calculated by binning the original $\overline{g+r}~$ ICL map by $4\times 4$ pixels to reduce the noise and then resized to the native size. The contours remark the clumpiness and filamentous morphology of the ICL, whose distribution is similar to that of the X-ray emitting gas (red contours), observed by SRG/eROSITA \citep{churazov2021}.\\

We confirm the detection of ICL clumps that are smaller than those described in Sect. \ref{sect:discussion:BGCs}, some of them already discovered in previous works \citep{adami2005a,jimenez-teja2019}. For example, we observe two clear clumps of red ICL in the southeastern part of the main cloud, around galaxies NGC 4921 and NGC 4911 (see Fig. \ref{fig:color:ICL}; these galaxies are marked with a blue square and triangle, respectively). This is very likely pre-processed ICL, still bound to the small groups that are identified around these two bright galaxies (groups \#41 and \#22). {Both groups have very different mean velocities, $6149\pm 938$ and $8051\pm 223$ km/s, respectively, so they are identified as independent structures. These velocities are higher than those found by \cite{adami2005b} ($5614\pm 159$ and $7627\pm 219$ km/s, respectively), but consistent considering the velocity dispersions}. Group \#22 has a lower velocity dispersion than \#41 and a more compact ICL with a faint tail extending toward the east (see Fig. \ref{fig:ICL+Xray}). This suggests that group \#22 is still quite bound and it could be infalling into Coma in a tangential trajectory from the east (and, possibly, from the back, given its mean velocity). Galaxy NGC 4911 was identified as an H$_{\alpha}$ emitter by \cite{iglesias2002}, which could be due to an enhanced star formation originated by the interaction of the group with Coma's intracluster medium. Contrarily, group \#41 has a very high velocity dispersion, its ICL is more spread and blue in the outskirts (Fig. \ref{fig:color:ICL}), which indicates that this group could have already passed tangentially by the core of Coma. Indeed, NGC 4921 is clearly displaced which respect to the mean velocity of the group (by $\sim -678$ km/s) and also spatially with respect to the rest of galaxies of the group (see Fig. \ref{fig:DS+} right). Additionally, it also seems to have a very faint tail of diffuse light linking it to the main cloud of ICL (which is more visible in Fig. \ref{fig:ICL+Xray}). All these pieces of evidence together suggest that group \#41 could be traveling from the northwest and has already interacted with Coma.\\

A chain of identified groups are aligned with the intracluster filaments detected by \cite{hyeonghan2024} via weak lensing, as can be observed in Fig. \ref{fig:DS+} left. It is suggestive of the intracluster filaments being an accretion channel for the groups of galaxies in the comic filaments. For example, the north intracluster filament seems to be responsible for incorporating groups \#7, 25, 29, 31, and 32. The last two groups are very close to the core of the cluster. Both of them are small and compact groups with intragroup light still attached, so they are apparently infalling now following a relatively radial trajectory. \\

The western filament, as named by \cite{hyeonghan2024}, which connects Coma with the cluster Abell 1367 (which is also part of the Coma supercluster) and the filamentous overdensity that extends from the core of Coma to the northwest in the approximate direction of the cluster Abell 779, are also important pathways for infalling groups. It is especially notorious the injection of preprocessed ICL from group \#20, northwest to the BCG NGG 4874, which is also associated with an H$_{\alpha}$-emitting galaxy \citep{iglesias2002}.\\

\subsection{NGC 4839 group} \label{sect:discussion:NGC4839}
Many works in the literature have discussed the dynamics of the group associated with NGC 4839 \citep[see. e.g., ][ and references therein]{oh2023}. While it is longtime known that this group is infalling into Coma, it is still widely debated whether the group is in a premerger or postmerger state. Although the most recent evidence supports the scenario where the NGC 4839 group is starting its second infall after crossing the core of Coma and reaching the apocenter, some earlier works found that the group may be in its first infall, mostly based on the dynamics of the Coma galaxies and the X-ray distribution of the gas as observed by ROSAT and XMM-Newton \citep[e.g., ][]{briel1992,white1993,colless1996,neumann2001,neumann2003}. The reprocessing of the XMM-Newton data, along with new observations from Chandra and SRG/eRosita, led to discover new substructures and features that were better explained within a postmerger scenario. A putative gas bridge between central region of Coma and NGC 4839, an envelope of hotter-than-the-outskirts gas surrounding the core of the NGC 4839 group, and an elongated tail of ram-pressured stripped gas spreading toward the opposite direction of the Coma center are solid evidences that support the postmerger hypothesis \citep{lyskova2019,churazov2021,churazov2023}. Two radio haloes that extend over the cluster core and the NGC 4839 group also appear connected, where a low surface brightness radio bridge is detected \citep{bonafede2021,lal2022}. Moreover, at the opposite direction of NGC 4839, toward the southwest, a brighter, filamentous radio emission is identified, linking this group to a radio relic, which is suggestive of diffusive shock-induced cosmic ray electron re-acceleration that could be originated by the merger driven runaway shock \citep{zhang2019}. The high compactness of the distribution of globular clusters around NGC 4839, especially evident when it is compared to that of the similar, neighboring galaxy NGC 4816, is compelling evidence that NGC 4839 may have lost the globular clusters from its outskirts during a first core passage \citep{oh2023}. The excess of poststarbust galaxies and the deficiency of HI emission in the NGC 4839 group can be also explained if the group already crossed the core of Coma and had its gas partially removed \citep{colless1996,healy2021}. It is worth mentioning though, that an alternate hypothesis to explain the depletion of HI is that NCG 4839 could have been accreted through the filament that connects Coma with Abell 1367 and this dense environment could have triggered the quenching of the star formation. \\

\begin{figure*}
\centering
\includegraphics[width=\textwidth]{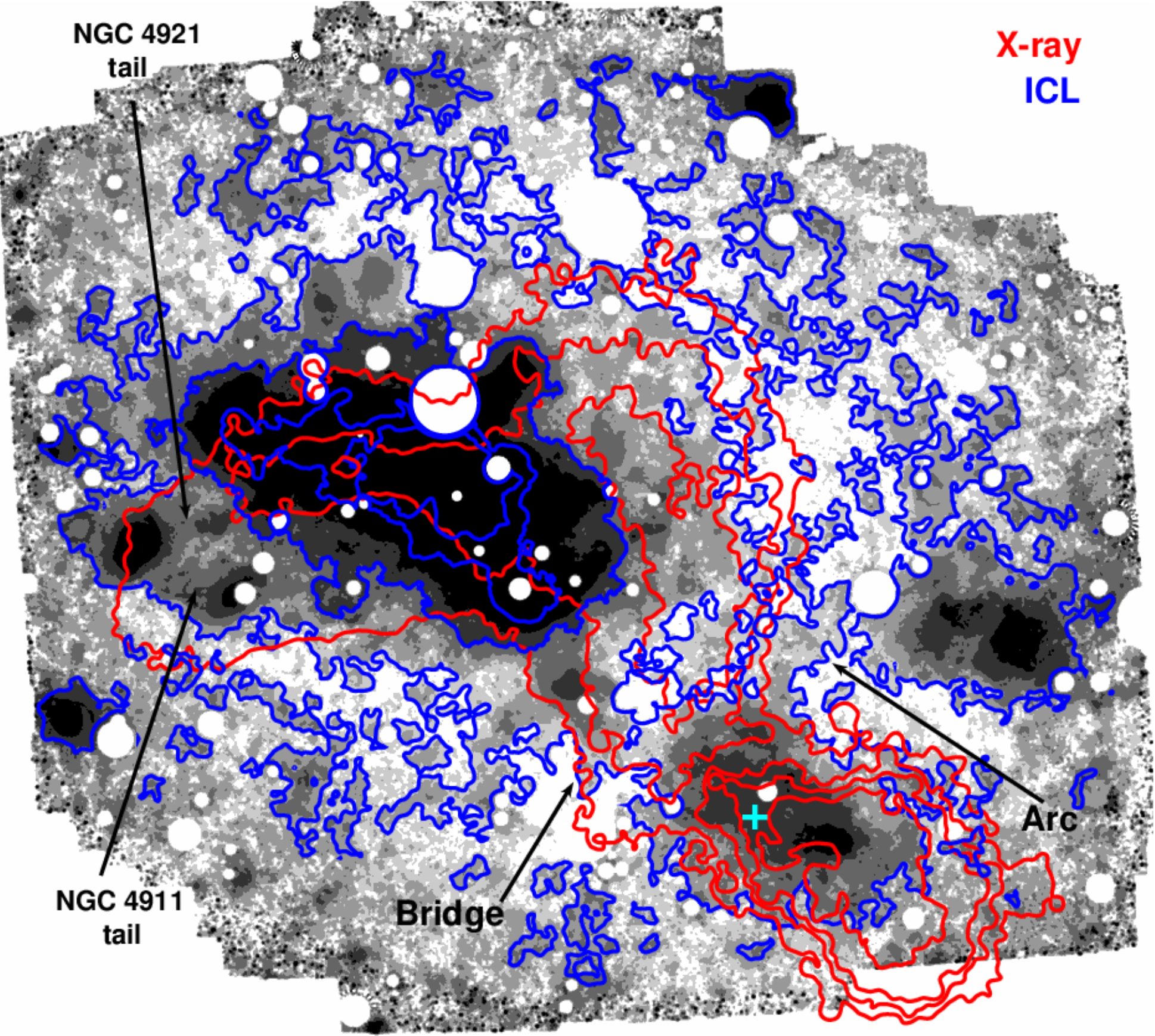}
\caption{Composite $\overline{g+r}~$ ICL map, threshold segmented to facilitate the visualization of the ICL. The main contours of the ICL (blue) and X-ray (red) distributions are overplotted. The ICL contours were computed from the original ICL map, just applying a binning of $4\times 4$ pixels and later resizing them to the native dimensions. The X-ray contours are derived from the SRG/eROSITA data and are a courtesy of \cite{churazov2021} (see their Figs. 6 and 11). We removed some of the smaller-scale contours for the sake of clarity. We highlight the ICL faint bridge linking the ICL clouds associated with the NGC 4839 group and the cluster core, which matches in projection the faint bridge discovered by \cite{churazov2021}. The cyan cross indicates a blue patch of ICL depicted in Fig. \ref{fig:color:ICL} and described in Sect. \ref{sect:discussion:NGC4839}. A faint and narrow arc of ICL connects NGC 4839 and NGC 4816, without X-ray emission detected, likely the remnant of an infalling galaxy located at the end of the arc.  North is up, east is left.}  
\label{fig:ICL+Xray}
\end{figure*}

In the pseudo-color Fig. \ref{fig:color:ICL}, we can observe that the ICL is distributed asymmetrically within the NGC 4839 group (this galaxy is marked with a blue cross). The ICL is more concentrated in the southwestern region of the group, with some arms and loops that connect the members of the group with the main cloud of intragroup light. This is probably intragroup light that is still bound to the NGC 4839 group and is in the process of being detached and ingested by the ICL of Coma. Indeed, the southwestern part of the NGC 4839 intragroup light is redder and sharper than the northeastern region, which is bluer, more diffuse, and extended in a fan-like shape (${\Delta (g-r)=0.35}$). If the NGC 4839 group was infalling for the first time into Coma from the western filament that feeds it, this ICL morphology would be hard to explain. However, if NGC 4839 followed a trajectory coming from the north, passing by the Coma BCGs and continuing its orbit until it reaches its apocenter and starts falling back again to the core of the cluster, as suggested by \cite{lyskova2019} and \cite{churazov2021}, it would leave a trail of stripped, blue ICL stars behind. The bluer color of the ICL in the northeast half of the group is explained by being mainly composed of stars thrown from the outskirts of galaxy NGC 4839, whereas the southwestern region would mostly contain preprocessed intragroup light, hence redder. This scenario is further supported by the presence of a faint bridge of ICL that links the NGC 4839 group with the main cloud of ICL that surrounds the cluster core, which is spatially coincident with a faint bridge of X-ray emission discovered by \cite{churazov2021} (see Fig. \ref{fig:ICL+Xray}). This bridge also matches the low surface brightness bridge of radio emission found in previous works \citep{bonafede2021,lal2022}.\\

\begin{figure}
\centering
\includegraphics[width=0.5\textwidth]{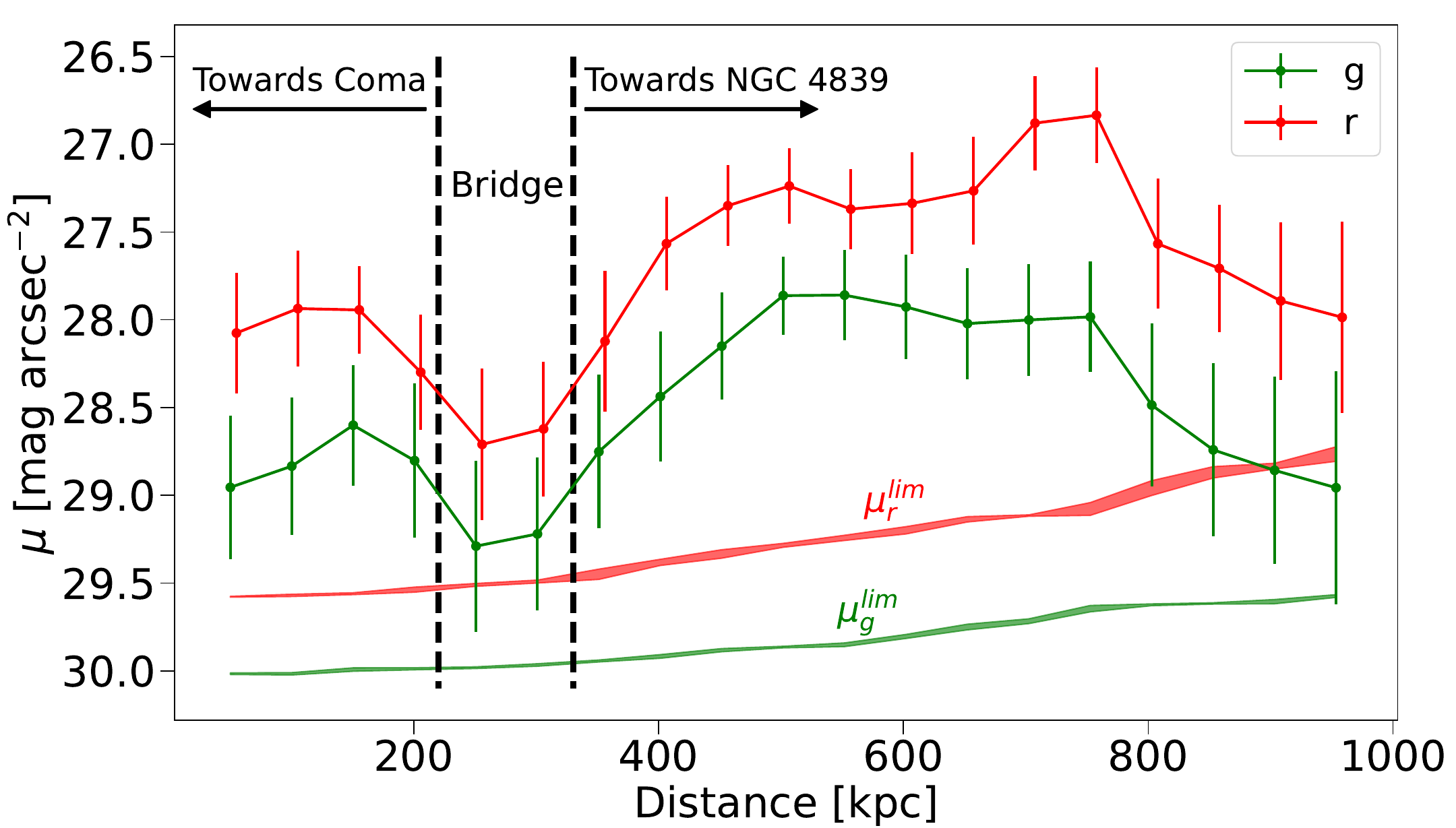}
\caption{Surface brightness profile along the bridge. Measurements are made in boxes parallel to the main axis, in the $g$ (green) and $r$ (red) bands. The green and red areas at the bottom indicate the limiting surface brightness computed from the maps elaborated by Román et al. (in prep), averaged inside the same boxes. The position of the bridge is marked with dashed vertical lines, and the directions towards the core of Coma and the NGC 4839 group are indicated with arrows.}  
\label{fig:bridge_profile}
\end{figure}

In Fig. \ref{fig:bridge_profile}, we show the ICL brightness profile along the bridge. The brightness is averaged inside contiguous boxes whose main axis is perpendicular to the bridge main axis. The boxes are $\sim 50\times 50$ kpc, {in order to have the same width as the boxes used by \cite{bonafede2021} (see their Fig. 3) to compute the profile of the radio bridge detected in the same region}. The same boxes are used to compute the average limiting surface brightness levels, using the maps provided by Román et al. (in prep). The average surface brightness of the bridge is $\sim 28.5$ and $29.2$ mag arcsec$^{-2}$ {in the $g$ and $r$ bands, respectively}, well above the limiting surface brightness of the region even considering the errorbars. {This corresponds to a correlated significance of $2.0\sigma$ and $2.1\sigma$, respectively}. This is the first time that the stellar distribution of the NGC 4839 group is observed to be connected with that of the main core of Coma, confirming that both structures are not detached and must have already interacted in a recent past. Indeed, our $g$ band traces the presence of the IE (see Sect. \ref{sect:discussion:dynamical_state}), which is very likely generated by an injection of A- and F-type stars into the ICL from the merging event \citep[see, e.g., ][]{jimenez-teja2018}. Similarly to \cite{jimenez-teja2019}, if we assume that these stars were stripped halfway through their lifetimes and also assuming an average age for these stellar types of $\sim 2.3$ Gyr using a Salpeter initial mass function, then NGC 4839 must have crossed the core of Coma less than $\sim 1.2$ Gyr ago. This is in excellent agreement with the calculations by \cite{oh2023}, who estimated this first crossing would have occurred $\sim 1.6$ Gyr ago from the analysis of the globular cluster distribution in the NGC 4839 group.\\

We also observe in Fig. \ref{fig:color:ICL} a very blue region of ICL (${g-r=0.18}$) close to the nucleus of NGC 4839, possibly the bluest patch of ICL in our field of view. This region is marked with a cyan cross in Fig. \ref{fig:ICL+Xray}, where we see that its western edge coincides with a concave X-ray contour. The return of NGC 4839 group after reaching the apocenter may cause the compression of the gas still attached to group, possibly triggering the star formation and being partially responsible for this blue ICL color. A further, more local analysis of this and other regions will be the subject of a future paper.\\

Finally, we can also observe in Fig. \ref{fig:ICL+Xray} an extremely faint, narrow arc of diffuse light connecting the NGC 4839 and NGC 4816 groups. However, the absence of X-ray emission and tidal features in the NGC 4816 group that could be indicative of merging activity would rule out an interaction between these two groups at first guess, although it is possible to have faint X-ray emission below the eROSITA's detection limit or even stripping by the intracluster medium without X-ray emission. We propose an alternate scenario instead, that this arc could be the remnant of a shredded galaxy, hypothesis that will be further explored in Román et al. (in prep). If we exclude this arc of our considerations, the intragroup light associated with NGC 4816 is still detached from the main cloud of ICL, compelling evidence that this group is still in a pre-merger state, in agreement with \cite{healy2021} findings.\\


\section{Conclusions} \label{sect:conclusions}

{We obtained extremely deep low surface brightness images of the Coma cluster taken as part of the HERON cluster project, in the $g$ and $r$ bands. We employed our data to analyze the projected distribution of the ICL and to connect its morphology and color with the dynamics of the brightest galaxies in the cluster. To identify the cluster member candidates, we used a large sample of spectroscopic measurements, gathered from the DESI EDR, the SDSS, and the NED. The main conclusions of this study are listed below:}
\begin{itemize}
\item We obtained the largest image of the ICL in an individual cluster to date, reaching radii as far as 1.5 Mpc from the cluster core. The ICL distribution is clumpy and not homogeneous, rich with tidal features and substructures that reflect the active dynamical processes that are ongoing in the cluster.
\item We identified a total of 2,157 galaxy member candidates in a region of $\sim 10.4\times 5.3$ Mpc around the center of Coma using the caustic technique.
\item We applied the DS+ algorithm to identify 42 possible groups in Coma. Many of these groups are located in or near the cluster core, while many others show a tendency to be located, in projection, in intracluster filaments previously detected via their weak lensing signal.
\item The ICL fractions of $19.9\pm 0.5\%$ and $19.6\pm 0.6\%$ in the $g$ and $r$ band respectively, confirm the active dynamical stage of Coma, in consonance with previous findings \citep{jimenez-teja2019}.
\item The region surrounding the two BCGs has the brightest ICL, with numerous clumps that can be associated with groups of galaxies. Both BCGs have several groups that have similar mean velocities, suggesting that they formed a larger structure in the past and may be in the process of dissolving and fragmenting due to their dynamical interaction after several core passages.
\item A handful of groups seems to be infalling into the main cloud of ICL, with tails of shredded intragroup light that can trace back their possible trajectories. The groups associated with the galaxies NGC 4921 and NGC 4911, southeast of the BCGs, are noteworthy examples. Their difference in velocity dispersion, direction of the intragroup light tails, intragroup light color, and distribution of galaxy members suggest that the two groups are independent structures in different dynamical stages, pre-merging and in an ongoing merger.
\item Similarly, the two groups associated with the bright ellipticals southwest of the Coma core, NGC 4839 and NGC 4816, show very different characteristics. NGC 4839 has an evolved intragroup light, more concentrated in the region where the group is believed to reach its apocenter and more diffuse and extended in the opposite side, perhaps indicating that it is leaving a trail of ICL stars after its core passage from the north. To reinforce this scenario, we discovered a faint bridge of ICL connecting Coma's main ICL body and the intragroup light of NGC 4839 also visible in the X-ray and optical counterparts. Contrarily, the intragroup light of NGC 4816 does not present any evidence of interaction with the Coma core, while being completely detached from the main cloud of ICL.
\end{itemize}

{This work reveals the potential of the ICL as a tracer of the dynamical stage and the recent dynamical history of galaxies in a cluster. The ICL, as the remnant material that is spread into the cluster potential after galaxies' encounters and interactions, keeps the record from which we can potentially unveil the recent history of the cluster's evolutionary past. }


\begin{acknowledgements}
Y.J-T  and J.M.V. acknowledge financial support from the Spanish MINECO grant PID2022-136598NB-C32 and from the State Agency for Research of the Spanish MCIU through the Center of Excellence Severo Ochoa award to the Instituto de Astrofísica de Andalucía (SEV-2017-0709) and grant CEX2021-001131-S funded by MCIN/AEI/ 10.13039/501100011033. Y.J-T. also acknowledges financial support from the European Union’s Horizon 2020 research and innovation programme under the Marie Skłodowska-Curie grant agreement No 898633 and the MSCA IF Extensions Program of the Spanish National Research Council (CSIC). J.R. acknowledges financial support from the Spanish Ministry of Science
and Innovation through the project PID2022-138896NB-C55. K. H. acknowledges support for the current research from the National Research Foundation (NRF) of Korea under the programmes 2022R1A2C1003130 and RS-2023-00219959. R.A.D. acknowledges partial support from CNPq grant 312565/2022-4. P.A.A.L. thanks the support of CNPq (grant 312460/2021-0) and FAPERJ (grant E-26/200.545/2023). R.M.R. acknowledges technical support for the Jeanne Rich Telescope from David Gedalia and Francis Longstaff of the POA, and acknowledges financial support from his late father, Jay Baum Rich. O.C. acknowledges support from a McNair Fellowship.\\

This research used data obtained with the Dark Energy Spectroscopic Instrument (DESI). DESI construction and operations is managed by the Lawrence Berkeley National Laboratory. This material is based upon work supported by the U.S. Department of Energy, Office of Science, Office of High-Energy Physics, under Contract No. DE–AC02–05CH11231, and by the National Energy Research Scientific Computing Center, a DOE Office of Science User Facility under the same contract. Additional support for DESI was provided by the U.S. National Science Foundation (NSF), Division of Astronomical Sciences under Contract No. AST-0950945 to the NSF’s National Optical-Infrared Astronomy Research Laboratory; the Science and Technology Facilities Council of the United Kingdom; the Gordon and Betty Moore Foundation; the Heising-Simons Foundation; the French Alternative Energies and Atomic Energy Commission (CEA); the National Council of Science and Technology of Mexico (CONACYT); the Ministry of Science and Innovation of Spain (MICINN), and by the DESI Member Institutions: www.desi.lbl.gov/collaborating-institutions. The DESI collaboration is honored to be permitted to conduct scientific research on Iolkam Du’ag (Kitt Peak), a mountain with particular significance to the Tohono O’odham Nation. Any opinions, findings, and conclusions or recommendations expressed in this material are those of the author(s) and do not necessarily reflect the views of the U.S. National Science Foundation, the U.S. Department of Energy, or any of the listed funding agencies.\\

Funding for the Sloan Digital Sky Survey IV has been provided by the Alfred P. Sloan Foundation, the U.S. Department of Energy Office of Science, and the Participating Institutions. SDSS acknowledges support and resources from the Center for High-Performance Computing at the University of Utah. The SDSS web site is www.sdss4.org.

SDSS is managed by the Astrophysical Research Consortium for the Participating Institutions of the SDSS Collaboration including the Brazilian Participation Group, the Carnegie Institution for Science, Carnegie Mellon University, Center for Astrophysics | Harvard \& Smithsonian (CfA), the Chilean Participation Group, the French Participation Group, Instituto de Astrofísica de Canarias, The Johns Hopkins University, Kavli Institute for the Physics and Mathematics of the Universe (IPMU) / University of Tokyo, the Korean Participation Group, Lawrence Berkeley National Laboratory, Leibniz Institut für Astrophysik Potsdam (AIP), Max-Planck-Institut für Astronomie (MPIA Heidelberg), Max-Planck-Institut für Astrophysik (MPA Garching), Max-Planck-Institut für Extraterrestrische Physik (MPE), National Astronomical Observatories of China, New Mexico State University, New York University, University of Notre Dame, Observatório Nacional / MCTI, The Ohio State University, Pennsylvania State University, Shanghai Astronomical Observatory, United Kingdom Participation Group, Universidad Nacional Autónoma de México, University of Arizona, University of Colorado Boulder, University of Oxford, University of Portsmouth, University of Utah, University of Virginia, University of Washington, University of Wisconsin, Vanderbilt University, and Yale University.
\end{acknowledgements}

%
%

\bibliographystyle{aa}
\bibliography{YJTbibliography}{}

\begin{appendix}
\section{Coma catalog} \label{appendix:catalog}
We list in Table \ref{table:members} the cluster member galaxies identified by the caustic technique (see Sect. \ref{sect:analysis:cluster_membership}). 

\begin{table}[h!]
\caption{Catalog of Coma galaxies within a radius of $\sim 3.8$ Mpc around the cluster center.}
\centering
\begin{tabular}{ccc}
R.A. & Dec & z\\ 
\hline
195.482649 & 29.323511 & 0.02379 \\
193.791277 & 26.005201 & 0.02548 \\
193.195419 & 26.342686 & 0.02184 \\
193.323356 & 26.339289 & 0.02312 \\
193.139149 & 26.470082 & 0.02389 \\
193.234722 & 26.594886 & 0.02430 \\
193.243189 & 26.400579 & 0.01981 \\
193.263145 & 26.476549 & 0.02566 \\
193.371211 & 26.494159 & 0.02084 \\
193.413720 & 26.475759 & 0.02079 \\
\hline
\end{tabular} 
\tablefoot{The full table is available as supplementary material}\label{table:members}
\end{table}
\end{appendix}

\end{document}